\documentclass[10pt,a4paper,twoside]{article}
\usepackage{epsfig}
\usepackage{baltlat6}
\usepackage{array}
\usepackage{here}
\begin{document}
\ \
\vspace{0.5mm}
\setcounter{page}{277}
\vspace{8mm}

\titlehead{Baltic Astronomy, vol.\,17, 277--282, 2008}

\titleb{2MASS TWO-COLOR INTERSTELLAR REDDENING LINES:\\
THE BAND-WIDTH EFFECT}

\begin{authorl}
\authorb{V. Strai\v{z}ys}{1} and
\authorb{Romualda Lazauskait\.e}{1,2}
\end{authorl}

\begin{addressl}
\addressb{1}{Institute of Theoretical Physics and Astronomy, Vilnius
University,\\  Go\v{s}tauto 12, Vilnius LT-01108, Lithuania;
straizys@itpa.lt}
\addressb{2}{Department of Theoretical Physics, Pedagogical University,
Student\c{u} 39,\\ Vilnius, LT-08106, Lithuania; laroma@itpa.lt}
\end{addressl}

\submitb{Received: 2008 December 2; accepted: 2008 December 15}

\begin{summary} The band-width effect on interstellar reddening lines in
the \hbox{$J$--$H$} vs.~$H$--$K_s$ diagram of the 2MASS survey is
investigated using synthetic color indices and color excesses based on
the Kurucz model atmospheres.  At large interstellar reddenings
($E_{H-K_s}$\,$\geq$\,1.0) reddening lines deviate considerably from a
straight line.  The lines can be approximated by a parabolic equation:
$E_{J-H} = r E_{H-K_s} + s E_{H-K_s}^2$ where the slope coefficient,
$r$, and the curvature coefficient, $s$, depend slightly on the
intrinsic energy distribution of the source.  The curvature of the
reddening lines is confirmed by the $J$--$H$ vs.~$H$--$K_s$ diagrams
plotted by Strai\v{z}ys and Laugalys (2008) from 2MASS observations.
\end{summary}

\begin{keywords} ISM:  extinction -- stars:  fundamental
parameters -- photometric systems: infrared, 2MASS \end{keywords}

\resthead{2MASS two-color interstellar reddening lines: the band-width
effect}
{V. Strai\v{z}ys, R. Lazauskait\.e}

\sectionb{1}{INTRODUCTION}

The color excess in a two-color monochromatic system (defined by the
wavelengths $\lambda_1$ and $\lambda_2$) is a difference of interstellar
extinctions $A(\lambda_1)$ and $A(\lambda_2)$ expressed in stellar
magnitudes.  The values of the monochromatic extinctions for the unit
dust mass $x$ can be taken from the interstellar extinction law, i.e.,
the dependence of $A$ on $\lambda$ or $\lambda^{-1}$.  In the
monochromatic or narrow-band photometric systems the extinction
increases linearly with increasing of the dust mass.

In the case of a heterochromatic photometric system the
extinctions are defined by the equation
$$
A_m = -2.5 \log {
{\int {F(\lambda)~R_{m}(\lambda)~\tau^x(\lambda)~d\lambda}}
\over
{\hstrut\int {F(\lambda)~R_{m}(\lambda)~d\lambda}}}~, \eqno(1)
$$
where $F(\lambda)$ is the spectral energy distribution function of
a star
or a model atmosphere, $R_m(\lambda)$ is the response function of the
passband, $\tau(\lambda)$ is the transmittance function of the unit
mass of dust and $x$ is the number of dust masses.

This means that the heterochromatic extinction depends on the spectral
energy distribution and the amount of interstellar dust.  A red star,
affected by the same cloud of interstellar dust, will exhibit smaller
extinction $A(\lambda)$ than a blue star.  Also, if a dust cloud gives
the extinction $A(\lambda)$, the addition of the second identical cloud
will raise the extinction not to up 2$A(\lambda)$ but to a smaller
quantity.  The broader the response function, the larger is the
dependence of the extinction on the spectral energy distribution and the
amount of interstellar reddening.  This dependence is known as the
band-width effect.  The reason for the effect can be understood as the
dependence of the effective wavelength on spectral type and interstellar
reddening.

Since color excesses are differences of extinctions in two passbands,
the dependence of $A(\lambda_1)$ and $A(\lambda_2)$ on spectral energy
distribution of the star and on its interstellar reddening transfers the
band-width effect to color excesses and color-excess ratios.  However,
in an exceptional case the band-width effect on a color excess can be
zero, when the band-width effect in both passbands is the same.

The band-width effect was well known to stellar photometrists long ago;
see, e.g., the reviews by one of the authors (Strai\v zys 1977,
1992). However, in some new photometric systems the effect sometimes
becomes forgotten. The near-infrared {\it J, H, K} system is one of such
examples.

Jones \& Hyland (1980) were probably the first who tried to
estimate the band-width effect on the form of reddening line in the
$J$--$H$ vs.~$H$--$K$ diagram.  By synthetic photometry they found some
deviation of heavily reddened stars at \hbox{$J$--$H$} $>$\,3.5.  A
similar
effect was also calculated by Nagata et al.  (1993).  Naoi et al.
(2006) found the decline of the reddening line slope in Ophiuchus and
Chamaeleon star-forming regions by observations in the SIRIUS {\it J,
H}, $K_s$ system, but failed to confirm the effect by synthetic
photometry.

One of the authors of the present paper (Strai\v zys 1992) has estimated
the band-width effect in the {\it UBVRIJHKLM} system by calculating
color excesses and their ratios for black bodies of different
temperatures.  A clear decline of the ratio $E_{J-H}/E_{H-K}$ from 2.0
to 1.7 was found when the temperature of the radiation source has
decreased from 20\,000 K to 2000 K.

Recently, during the investigation of the $E_{J-H}/E_{H-K_s}$ ratios in
various Milky Way directions and in star-forming regions (Strai\v zys \&
Laugalys 2008), we have noted that in most directions heavily reddened
stars deviate down from the linear reddening line of red giants.  This
stimulated the investigation of possible band-width effect for heavily
reddened stars in the two-color diagram of the 2MASS system.

\sectionb{2}{CALCULATIONS AND RESULTS}

Interstellar extinctions in the passbands of the 2MASS system were
calculated by Equation (1) with the functions taken from the following
sources.  Spectral energy distributions $F(\lambda)$ were taken for 409
synthetic spectra of solar metallicity and various temperatures and
gravities from Kurucz (2001).  Response functions of the 2MASS passbands
were taken from Cutri et al.  (2006) and Skrutskie et al.  (2006).  The
transmittance function of the interstellar dust for a unit mass ($x$ =
1, this corresponds to $E_{B-V}$ = 1.0) is taken from Strai\v zys (1992,
Table 3), with some small modification at wavelengths longer than 2.0
$\mu$m to adjust the extinction law to the ratio of color excesses
$E_{J-H}/E_{H-K_s}$ = 1.9.  In calculations the dust mass $x$ was varied
from 2 to 10; these values correspond to $A_V$ = 6.2 and 31 mag.

In Table 1 we present the calculated color excesses and their ratios for
a selected set of 87 models with different temperatures and gravities
and for five values of $x$ to show the significance of the band-width
effect.  For the model with $T_{\rm eff}$ = 35\,000 K, which corresponds
to the spectral class O8, the ratio of color excesses is 1.99 at $x$ = 2
and 1.85 at $x$ = 10.  For the model with $T_{\rm eff}$ = 4500 K and
$\log g$ = 2.5, which corresponds to red clump giants (K2\,III), the
ratio is 1.95 for $x$ = 2 and 1.81 for $x$ = 10.

\begin{table}[!t]
\begin{center}
\vbox{\footnotesize\tabcolsep=3pt
\parbox[c]{124mm}{\baselineskip=10pt
{\smallbf\ \ Table 1.}{\small\
Ratios of color excesses $E_{J-H}/E_{H-K_s}$ for the Kurucz models
with various interstellar extinctions.\lstrut}}
\begin{tabular}{cccccc|cccccc}
\hline
 $T_{\rm eff}$, $\log g$ & $x$\,=\,2 & $x$\,=\,4 & $x$\,=\,6 & $x$\,=\,8 & $x$\,=\,10 &
  $T_{\rm eff}$, $\log g$ & $x$\,=\,2 & $x$\,=\,4 & $x$\,=\,6 & $x$\,=\,8 & $x$\,=\,10 \hstrut\lstrut\\
\hline
 3500, 1.0 & 1.942  &  1.906  &  1.871  &  1.838  &  1.808  &    7500, 5.0 & 1.972  &  1.930  &  1.894  &  1.862  &  1.830 \hstrut \\
 3500, 2.0 & 1.939  &  1.901  &  1.867  &  1.836  &  1.806  &    8000, 1.0 & 1.989  &  1.954  &  1.916  &  1.882  &  1.850         \\
 3500, 3.0 & 1.942  &  1.905  &  1.868  &  1.836  &  1.807  &    8000, 2.0 & 2.000  &  1.958  &  1.921  &  1.887  &  1.853         \\
 3500, 4.0 & 1.943  &  1.903  &  1.870  &  1.837  &  1.807  &    8000, 3.0 & 1.996  &  1.954  &  1.916  &  1.883  &  1.850         \\
 3500, 5.0 & 1.940  &  1.902  &  1.868  &  1.834  &  1.804  &    8000, 4.0 & 1.982  &  1.945  &  1.910  &  1.874  &  1.844         \\
 4000, 1.0 & 1.932  &  1.900  &  1.864  &  1.832  &  1.803  &    8000, 5.0 & 1.975  &  1.937  &  1.900  &  1.866  &  1.835         \\
 4000, 2.0 & 1.939  &  1.898  &  1.865  &  1.832  &  1.802  &    8500, 2.0 & 2.004  &  1.961  &  1.922  &  1.888  &  1.855         \\
 4000, 3.0 & 1.936  &  1.898  &  1.864  &  1.831  &  1.801  &    8500, 3.0 & 2.000  &  1.959  &  1.920  &  1.887  &  1.854         \\
 4000, 4.0 & 1.939  &  1.900  &  1.864  &  1.831  &  1.802  &    8500, 4.0 & 1.989  &  1.952  &  1.915  &  1.881  &  1.849         \\
 4000, 5.0 & 1.936  &  1.902  &  1.869  &  1.836  &  1.807  &    8500, 5.0 & 1.982  &  1.942  &  1.906  &  1.873  &  1.840         \\
 4500, 1.0 & 1.943  &  1.906  &  1.870  &  1.838  &  1.807  &    9000, 2.0 & 1.996  &  1.958  &  1.921  &  1.887  &  1.855         \\
 4500, 2.0 & 1.940  &  1.904  &  1.868  &  1.835  &  1.805  &    9000, 3.0 & 1.996  &  1.960  &  1.923  &  1.888  &  1.856         \\
 4500, 3.0 & 1.943  &  1.900  &  1.866  &  1.834  &  1.804  &    9000, 4.0 & 1.993  &  1.953  &  1.919  &  1.882  &  1.852         \\
 4500, 4.0 & 1.936  &  1.900  &  1.864  &  1.833  &  1.802  &    9000, 5.0 & 1.986  &  1.947  &  1.912  &  1.876  &  1.845         \\
 4500, 5.0 & 1.940  &  1.899  &  1.863  &  1.833  &  1.802  &    9500, 2.0 & 1.996  &  1.960  &  1.921  &  1.888  &  1.855         \\
 5000, 1.0 & 1.947  &  1.908  &  1.873  &  1.841  &  1.811  &    9500, 3.0 & 2.000  &  1.961  &  1.926  &  1.888  &  1.857         \\
 5000, 2.0 & 1.943  &  1.906  &  1.871  &  1.840  &  1.809  &    9500, 4.0 & 1.993  &  1.958  &  1.920  &  1.886  &  1.854         \\
 5000, 3.0 & 1.940  &  1.906  &  1.871  &  1.839  &  1.808  &    9500, 5.0 & 1.993  &  1.951  &  1.914  &  1.880  &  1.848         \\
 5000, 4.0 & 1.943  &  1.908  &  1.871  &  1.837  &  1.808  &   10000, 2.0 & 1.996  &  1.960  &  1.921  &  1.886  &  1.853         \\
 5000, 5.0 & 1.940  &  1.903  &  1.867  &  1.836  &  1.805  &   10000, 3.0 & 2.004  &  1.960  &  1.925  &  1.888  &  1.857         \\
 5500, 1.0 & 1.954  &  1.917  &  1.881  &  1.847  &  1.816  &   10000, 4.0 & 1.996  &  1.960  &  1.922  &  1.888  &  1.855         \\
 5500, 2.0 & 1.958  &  1.915  &  1.881  &  1.846  &  1.816  &   10000, 5.0 & 1.993  &  1.953  &  1.917  &  1.882  &  1.850         \\
 5500, 3.0 & 1.951  &  1.913  &  1.876  &  1.844  &  1.813  &   11000, 3.0 & 2.000  &  1.961  &  1.924  &  1.889  &  1.857         \\
 5500, 4.0 & 1.947  &  1.910  &  1.875  &  1.842  &  1.812  &   11000, 4.0 & 1.990  &  1.958  &  1.921  &  1.888  &  1.855         \\
 5500, 5.0 & 1.947  &  1.910  &  1.874  &  1.842  &  1.811  &   11000, 5.0 & 1.996  &  1.956  &  1.919  &  1.884  &  1.853         \\
 6000, 1.0 & 1.965  &  1.926  &  1.890  &  1.856  &  1.826  &   12000, 3.0 & 1.993  &  1.958  &  1.921  &  1.886  &  1.855         \\
 6000, 2.0 & 1.961  &  1.921  &  1.887  &  1.853  &  1.822  &   12000, 4.0 & 2.000  &  1.960  &  1.922  &  1.889  &  1.857         \\
 6000, 3.0 & 1.951  &  1.919  &  1.883  &  1.849  &  1.819  &   12000, 5.0 & 1.993  &  1.954  &  1.919  &  1.883  &  1.853         \\
 6000, 4.0 & 1.951  &  1.917  &  1.880  &  1.848  &  1.817  &   13000, 3.0 & 1.993  &  1.958  &  1.921  &  1.887  &  1.855         \\
 6000, 5.0 & 1.947  &  1.915  &  1.878  &  1.847  &  1.815  &   13000, 4.0 & 2.000  &  1.958  &  1.922  &  1.887  &  1.855         \\
 6500, 1.0 & 1.975  &  1.937  &  1.900  &  1.866  &  1.834  &   13000, 5.0 & 1.990  &  1.954  &  1.919  &  1.884  &  1.853         \\
 6500, 2.0 & 1.972  &  1.933  &  1.896  &  1.863  &  1.832  &   14000, 3.0 & 1.996  &  1.958  &  1.922  &  1.887  &  1.855         \\
 6500, 3.0 & 1.968  &  1.929  &  1.892  &  1.859  &  1.827  &   14000, 4.0 & 2.000  &  1.960  &  1.920  &  1.888  &  1.854         \\
 6500, 4.0 & 1.958  &  1.923  &  1.888  &  1.854  &  1.822  &   14000, 5.0 & 1.993  &  1.956  &  1.918  &  1.884  &  1.852         \\
 6500, 5.0 & 1.954  &  1.919  &  1.882  &  1.850  &  1.819  &   15000, 3.0 & 2.000  &  1.960  &  1.922  &  1.887  &  1.854         \\
 7000, 1.0 & 1.986  &  1.947  &  1.910  &  1.876  &  1.843  &   15000, 4.0 & 2.000  &  1.960  &  1.922  &  1.887  &  1.854         \\
 7000, 2.0 & 1.982  &  1.942  &  1.907  &  1.871  &  1.841  &   15000, 5.0 & 1.993  &  1.954  &  1.920  &  1.885  &  1.852         \\
 7000, 3.0 & 1.968  &  1.933  &  1.900  &  1.865  &  1.834  &   20000, 5.0 & 1.993  &  1.956  &  1.919  &  1.885  &  1.852         \\
 7000, 4.0 & 1.965  &  1.930  &  1.893  &  1.860  &  1.828  &   25000, 5.0 & 1.997  &  1.954  &  1.919  &  1.885  &  1.853         \\
 7000, 5.0 & 1.965  &  1.928  &  1.891  &  1.855  &  1.824  &   30000, 5.0 & 2.000  &  1.958  &  1.922  &  1.887  &  1.855         \\
 7500, 1.0 & 1.986  &  1.951  &  1.915  &  1.881  &  1.849  &   35000, 5.0 & 1.997  &  1.958  &  1.922  &  1.886  &  1.854         \\
 7500, 2.0 & 1.993  &  1.951  &  1.915  &  1.880  &  1.849  &   40000, 5.0 & 2.000  &  1.958  &  1.922  &  1.886  &  1.854         \\
 7500, 3.0 & 1.986  &  1.945  &  1.908  &  1.875  &  1.843  &   50000, 5.0 & 2.000  &  1.958  &  1.922  &  1.888  &  1.856         \\
 7500, 4.0 & 1.975  &  1.938  &  1.901  &  1.868  &  1.836  &              &        &         &         &         &        \lstrut \\
\hline
\end{tabular}
}
\end{center}
\vskip-8mm
\end{table}

In Figure 1 we plot the reddening line of red clump giants on the
$J$--$H$ vs.~$H$--$K_s$ diagram in a 1\degr\ diameter area in the
direction of $\ell$ = 330\degr, $b$ = 0\degr\ (Norma) taken from Strai\v
zys \& Laugalys (2008).  The theoretical line fits the observed points
very well.  In other Milky Way areas investigated by Strai\v zys \&
Laugalys (2008) the correspondence is not so good since the observed
reddening lines exhibit a slightly larger slope.  In most of the
star-forming regions investigated in that paper, the observed reddening
lines end at lower values of color indices and are not suitable for
verification of the reddening line curvature.


\begin{figure}[!tH]
\vbox{
\centerline{\psfig{figure=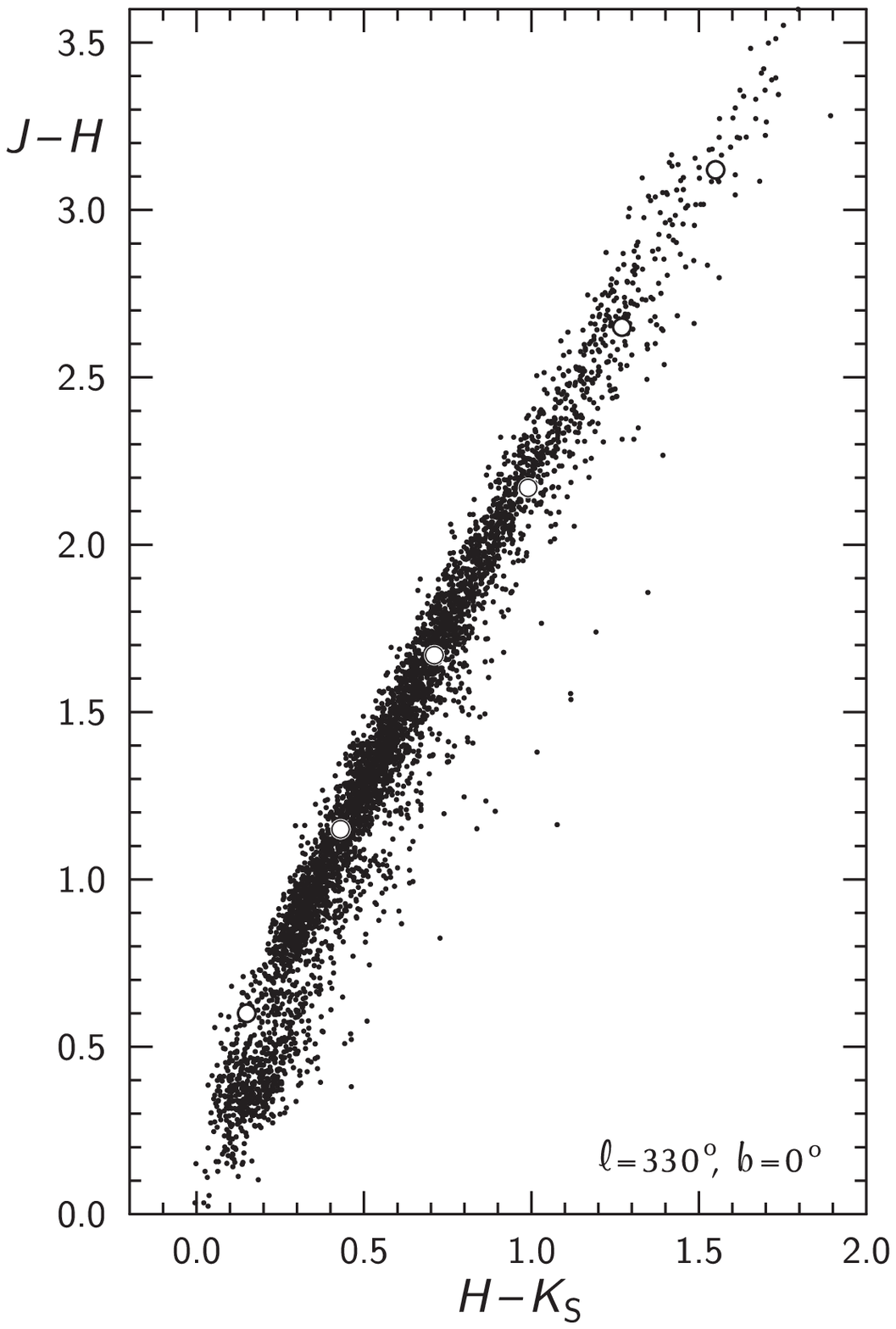,width=100mm,angle=0,clip=}}
\vspace{1mm}
\captionb{1}
{Synthetic reddening line for the Kurucz model $T_{\rm eff}$ =
4500 K, log\,$g$ = 4.0 (white circles) plotted on the observational
2MASS $J$--$H$ vs.~$H$--$K_s$ diagram in the direction with the Galactic
coordinates $\ell$ = 330\degr, $b$ = 0\degr.}
}
\end{figure}

The reddening line can be expressed by a parabolic equation
$$
E_{J-H}/E_{H-K_s} = r - s E_{H-K_s}. \eqno(2)
$$
The coefficient $r$ due to the band-width effect shows the
usual dependence on the temperature (or on spectral class), decreasing
from 2.03 for O-stars down to 1.96 for M-stars. The coefficient $s$
is almost constant, its average value is --0.12.

We also calculated effective wavelengths of the $J$, $H$ and $K_s$
passbands for various temperatures and gravities defined by the
following equation:
$$
\lambda_{\rm eff} = {\int
{F(\lambda)~R_{m}(\lambda)~\tau^{x}(\lambda)~\lambda~d\lambda} \over
\hstrut\int {F(\lambda)~R_{m}(\lambda)~\tau^{x}(\lambda)~d\lambda}}~.
\eqno(3)
$$
The results for five selected models of different
temperatures are listed in Table 2. The largest change of the effective
wavelengths both with the temperature and interstellar reddening is
observed for the $J$ passband:  0.01--0.02 $\mu$m between $T_{\rm eff}$
= 3500 K and 35000 K and 0.04--0.06 $\mu$m between $x$ = 0 and 10.  For
the $H$ passband the corresponding variations are 0.01 $\mu$m and 0.02
$\mu$m.  For the $K_s$ passband these variations are 0.002 $\mu$m and
0.018 $\mu$m.

The variations of $\lambda_{\rm eff}$ for the three passbands help to
understand why the reddening line in the $J$--$H$ vs.~$H$--$K_s$
diagram at large reddenings is curved down:  with increasing reddening
the shift of $\lambda_{\rm eff}$ for $J$ is much larger than for $H$ and
this leads to decrease of the base-line of the $J$--$H$ color.  As a
consequence, the increase of $J$--$H$ is slowed down in comparison with
the dust mass $x$.  At the same time, the difference of $\lambda_{\rm
eff}$ variation between the $H$ and $K_s$ passbands is much smaller, and
the values of $H$--$K_s$ color remain almost proportional to $x$ with
increasing reddening.

Since the effective wavelengths depend on the temperature and
reddening, their change should be taken into account when plotting the
interstellar extinction law:  the values of $A_{\lambda}$ determined for
early-type or less reddened stars should be plotted at shorter
wavelengths than for late-type or heavily reddened stars.

\begin{table}[!t]
\begin{center}
\vbox{\small\tabcolsep=6pt
\parbox[c]{124mm}{\baselineskip=10pt
{\normbf\ \ Table 2.}{\norm\
Effective wavelengths of the 2MASS passbands $J$, $H$ and $K_s$ in
$\mu$m
for Kurucz models of five values of temperatures and different values of
interstellar dust masses ($x$ = 0, 2, 4, 6, 8 and 10).\lstrut}}
\begin{tabular}{c|ccc|ccc}
\hline
$T_{\rm eff}$, log\,$g$ & $J$ & $H$ & $K_s$ & $J$ & $H$ & $K_s$\sstrut\\
\hline
         & \multicolumn{3}{c}{$x$ = 0} & \multicolumn{3}{c}{$x$ = 6}\sstrut \\
\hline
 3500, 4.0 &  1.253 &  1.644 &  2.145   &  1.290 &  1.656 &  2.156\hstrut \\
 4500, 4.0 &  1.250 &  1.640 &  2.144   &  1.288 &  1.653 &  2.155 \\
 6000, 4.0 &  1.244 &  1.637 &  2.144   &  1.283 &  1.651 &  2.155 \\
10000, 4.0 &  1.238 &  1.636 &  2.143   &  1.277 &  1.650 &  2.155 \\
35000, 4.0 &  1.236 &  1.634 &  2.143   &  1.275 &  1.648 &  2.154 \\
\hline
         & \multicolumn{3}{c}{$x$ = 2} & \multicolumn{3}{c}{$x$ = 8}\sstrut \\
\hline
 3500, 4.0 &  1.266 &  1.648 &  2.148   & 1.300 &  1.660 &  2.160 \hstrut\\
 4500, 4.0 &  1.264 &  1.645 &  2.147   & 1.299 &  1.657 &  2.158 \\
 6000, 4.0 &  1.258 &  1.642 &  2.147   & 1.294 &  1.655 &  2.159 \\
10000, 4.0 &  1.251 &  1.641 &  2.147   & 1.288 &  1.654 &  2.158 \\
35000, 4.0 &  1.249 &  1.639 &  2.146   & 1.287 &  1.652 &  2.158 \\
\hline
         & \multicolumn{3}{c}{$x$ = 4} & \multicolumn{3}{c}{$x$ = 10}\sstrut \\
\hline
 3500, 4.0 &  1.278 &  1.652 &  2.152   & 1.310 &  1.664 &  2.163 \hstrut\\
 4500, 4.0 &  1.276 &  1.649 &  2.151   & 1.308 &  1.661 &  2.162 \\
 6000, 4.0 &  1.271 &  1.646 &  2.151   & 1.304 &  1.659 &  2.162 \\
10000, 4.0 &  1.264 &  1.645 &  2.151   & 1.299 &  1.658 &  2.162 \\
35000, 4.0 &  1.263 &  1.643 &  2.150   & 1.297 &  1.656 &  2.161 \\
\hline
\end{tabular}
}
\end{center}
\vskip-4mm
\end{table}

\sectionb{3}{CONCLUSIONS}

Applying the method of synthetic photometry for the Kurucz models we
show that interstellar reddening lines in the 2MASS $J$--$H$
vs.~$H$--$K$ diagram due to the band-width effect are of parabolic form
with a curvature coefficient of $s$ = --0.12.  The slope of the
reddening line at constant reddening also decreases with decreasing
temperature, but this effect is much smaller.  The theoretical results
are confirmed by the observed reddening lines in the inner Galaxy
investigated by Strai\v zys \& Laugalys (2008).

The knowledge of the band-width effect in the {\it J, H}, $K_s$ system
on the slope and curvature of reddening lines, as well as on the
effective wavelengths, is important in determining the interstellar
extinction law in the infrared range (see, e.g., Fitzpatrick 1999;
Fitzpatrick \& Massa 2005, 2007; Indebetouw et al. 2005; Flaherty et al.
2007; Rom\'an-Z\'uniga et al. 2007).  If one accepts that the reddening
line is straight and solves all stars together, the ignorance of the
curvature can lead to a smaller ratio of color excesses.  Also, the
ignorance of the curvature of reddening lines can lead to wrong
classifications of heavily reddened stars from photometric data.

\References

\refb Cutri R. M., Skrutskie M. F., Van Dyk S., Beichman C. A. et al.
2006, Eplanatory Supplement to the 2MASS All Sky Data Release and
Extended Mission Products, \\
http://www.ipac.caltech.edu/2mass/releases/allsky/doc/explsup.html

\refb Fitzpatrick E. L. 1999, PASP, 111, 63

\refb Fitzpatrick E. L., Massa D. 2005, AJ, 130, 1127

\refb Fitzpatrick E. L., Massa D. 2007, ApJ, 663, 320

\refb Flaherty K. M., Pipher J. L., Megeath S. T. 2007, ApJ, 663, 1069

\refb Indebetouw R., Mathis J. S., Babler B. L. et al. 2005, ApJ, 619,
931

\refb Jones T. J., Hyland A. R. 1980, MNRAS, 192, 359

\refb Kurucz R. L. 2001,
http://kurucz.harvard.edu/grids/gridP00/fp00k2.pck

\refb Nagata T., Hyland A. R., Straw S. M. et al. 1993, ApJ, 406, 501

\refb Naoi T., Tamura M., Nakajima Y. et al. 2006, ApJ, 640, 373

\refb Rom\'an-Z\'uniga C. G., Lada C. J., Muench A., Alves J. F. 2007,
ApJ, 664, 357

\refb Skrutskie M. F., Cutri R. M., Stiening R., Weinberg M. D. et al.
2006, AJ, 131, 1163

\refb Strai\v{z}ys V. 1977, in {\it Multicolor Stellar Photometry},
Mokslas Publishers, Vilnius, Lithuania

\refb Strai\v{z}ys V. 1992, in {\it Multicolor Stellar Photometry},
Pachart Publishing House, Tucson, Arizona

\refb Strai\v{z}ys V., Laugalys V. 2008, Baltic Astronomy, 17, 253 (this
issue)

\end{document}